\documentclass[conference]{IEEEtran}
\IEEEoverridecommandlockouts
\usepackage{cite}
\usepackage{amsmath,amssymb,amsfonts}
\usepackage{algorithmic}
\usepackage{graphicx}
\usepackage{textcomp}
\usepackage{xcolor}
\usepackage{multirow}
\def\BibTeX{{\rm B\kern-.05em{\sc i\kern-.025em b}\kern-.08em
    T\kern-.1667em\lower.7ex\hbox{E}\kern-.125emX}}
\begin{document}

\title{Lightweight Encryption for the Low Powered IoT Devices\\
}

\author{\IEEEauthorblockN{Muhammad Usman}
    
        \IEEEauthorblockA{Department of Computer Engineering, Chosun University, Republic of Korea\\
        Email: usman@chousn.kr
    }

}

\maketitle

\begin{abstract}
The internet of things refers to the network of devices connected to the internet and can communicate with each other. The term things is to refer non-conventional devices that are usually not connected to the internet. The network of such devices or things is growing at an enormous rate. The security and privacy of the data flowing through these things is a major concern. The devices are low powered and the conventional encryption algorithms are not suitable to be employed on these devices. In this correspondence a survey of the contemporary lightweight encryption algorithms suitable for use in the IoT environment has been presented. 
\end{abstract}


\section{Introduction}
\IEEEPARstart{T}{he} internet of things makes use of the existing global internet infra-structure in order to connect the physical objects including devices, home appliances and vehicles to name a few. The traditional fields of embedded systems including the Radio-frequency identification (RFID), wireless sensor networks (WSN) have all contributed towards building up and empower the IoT. The IoT enables the objects and sensors to be identified uniquely on the network and there are billions of such devices interconnected to each other. According to an estimate, a revenue of more than $714$ billion will be generated by the year $2022$ with more than $30$ billion connected devices to the internet \cite{vermesan2013internet, okkonen2013internet}.  The IoT is focused towards creating smart environments and autonomous devices for use in constructing smart cities, smart transportation and smart hospitals \cite{miorandi2012internet, whitmore2015internet}. New IoT applications are introduced at a high pace, and the are generating huge amount of data that can have security and privacy threat. Billions of devices connected to each other shall pronounce the security related issues and shall expose the weaknesses in the infrastructure. It has been predicted that if the security measures are not taken into account, the attacks and malfunctions would outweigh the advantages of IoT applications \cite{roman2011securing}. The things in the IoT environment are constrained due to the limited battery life, size and computation, therefore, they cannot support the existing security solutions. The traditional cryptography algorithms and security schemes are insufficient for the ever evolving IoT applications which require scalable and robust solutions \cite{yang2010security,suo2012security,e2018fpga}. This requires the development of novel designs and architectures to effectively deal with the security and privacy issues suitable for the resource constrained devices in the IoT \cite{al2015internet,Ibrahim2017,7838585}. This correspondence is aimed to review the security and privacy challenges in the IoT and survey the contemporary lightweight cryptographic algorithms developed for the resource constrained devices. 
 
The rest of the paper is organized as follows: Section \ref{sec: BG} presents the security and privacy  challenges in the internet of things, a review of lightweight cryptography algorithms are presented in Section \ref{sec: LWC}  and the paper is concluded in Section \ref{sec: conclusion}.

\section{Security and Privacy Challenges in IoT} \label{sec: BG}
One of the aspect towards deployment of IoT at the root level is to ensure its security. The proposed security solution must consider the resource constrained devices i.e. they must be lightweight and should consume less memory and computation power. Some of the lightweight solutions presented recently does not take into consideration the inter-operability and integration into the IoT. The interconnected devices have varied computational capacity and operate on diversified protocols, thus the security models must operate on the global standards \cite{yang2017survey}. Another issue in the IoT is that of the privacy, which is the privilege of an individual or a device to define the extent to which their data will be presented or shared with other elements on the network. In the present situation where a large amount of data is generated and is available on the network, the individual must have appropriate control of their data that can be accessed by other entities. A framework for the privacy support is yet to be developed by the researchers that will provide scalability in the IoT. In the following section, the security and privacy challenges are explicitly discussed.

\subsection{Security in IoT}
The security refers to the protection of the devices and data in the IoT. Different layers of IoT can be protected by implying different security technologies. Some of the important security aspects in the IoT are authentication, confidentiality, integrity and availability. 

\subsubsection{Authentication}
Authentication is vital requirement in the IoT, as it is sorely required to keep the information away from the unauthorized devices and person. Traditionally, the internet users authenticate themselves using the Secure Socket Layers (SSL) through their browsers, where they present a secret key (passwords). However, the mechanism of the passwords in the automated IoT environment does not present a good solution. In the IoT, the devices are usually connected to a central hub or gateway. Since the gateway is responsible for the information exchange between the devices and cloud, the authentication can be implied at the gateway to ensure the information is transmitted to the authenticated device or user. By following proper authentication protocols, it can be guaranteed that the information shall not land into wrong hands. For the scenarios in the absence of gateway, where the sensors themselves are responsible for the information exchange to and from the cloud, lightweight authentication protocols are highly desirable \cite{fadi2020seamless, dammak2019token, walshe2019non}. 

\subsubsection{Confidentiality}
The exchange of the private information between the billions of IoT devices and their storage must possess confidentiality \cite{boltz2020context}. The components in the IoT collaborating with each other in order to provide the desired service, are vulnerable to the confidentiality attacks if an unauthorized access is made to the sensitive information. Confidentiality being a primary concern, can be ensured by employing an access control method or using a lightweight encryption scheme.

\subsubsection{Integrity}
Integrity refers to the protected data that is transmitted in its original form, without being altered by the cyber criminals. Error detection techniques such as cyclic redundancy check (CRC) can be utilized to ensure the uniqueness of the message. Furthermore, Secure Hash Algorithm (SHA) are also regularly used which is a mathematical algorithm to ensure the elements of the data are unchanged. Suitable frameworks to provide integrity in the IoT environment are thoroughly discussed in \cite{dhanvijay2019internet}.

\subsubsection{Availability}
In the IoT, the data must be available at all times to the authorized users. Back-end cloud and storage devices must present the data, software service and hardware whenever required. The software availability refers to the service that has been given to the user to access or modify the information, whereas the hardware availability refers to the availability of the gadgets and their access. 

\subsection{Privacy}
The objects or things in the IoT tends to communicate the data autonomously, therefore, the the information is vulnerable to attacks and threat. It is highly desirable that the information is kept safe during the autonomous transfer by ensuring the privacy. The end-to-end transfer of information is somewhat immune to the attacks, however, the communication made via variety of nodes and sensors is highly susceptible to privacy breach. Large amount of data being shared and collected in different IoT technologies, is often human centered. A person or the entity must verify the amount of information they are willing to share to others. In the later section we review some of the cryptography algorithms that are tailored for the resource constrained devices and gadgets utilized in the IoT. 

\section{Lightweight Cryptography for IoT}\label{sec: LWC}
Two types of cryptography algorithms exists; symmetric and asymmetric, they differ from each other on the basis of key that is used for encryption and decryption. The symmetric ciphers use same key for the encryption and decryption, on the other hand, the asymmetric ciphers makes use of a public and private key for the sake of encrypting and decrypting the data. The lightweight encryption algorithms commonly utilize the symmetric key algorithms since the large key sizes in the asymmetric algorithms are not suitable for the IoT. Another feature of using Feistel network in the symmetric key algorithms is that the encryption and decryption process are complement of each other, which reduces their code size and saves the memory and circuit of the constrained device and results in the reduced latency. The symmetric encryption algorithms however, can be use either block or stream methods to generate the cipher text. The block cipher accepts a fixed length block of input bits as the plain text and performs transformation according the key. They are usually composed of substitution and permutation network (SPN) which contains substitution boxes and permutation boxes to generate the cipher text from the plain text. Symmetric algorithms employing a Feistel network for the encryption of the data can utilize similar structure for encryption and decryption. Some of the lightweight encryption algorithm are discussed in the following. In the following, we shall  discuss about the recently proposed lightweight symmetric block ciphers that can be utilized in the IoT. 

\subsection{PRESENT}
\begin{figure}[ht]
	\begin{center}
		\includegraphics*[width=8cm]{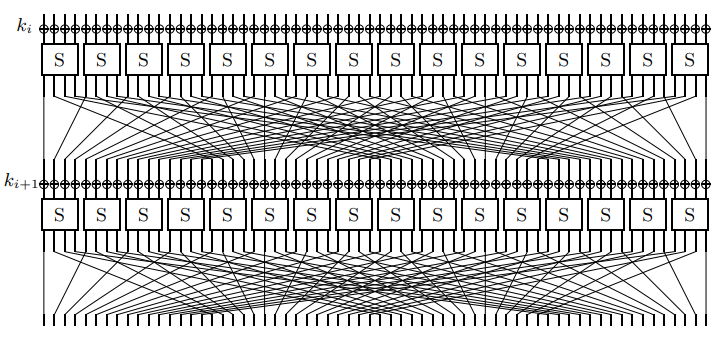}
	\end{center}
	\caption{S-P Network of PRESENT}
	\label{fig: SP_Present}
\end{figure}

PRESENT \cite{bogdanov2007present} was developed for the resource constrained devices having a block size of $80$ bits. The algorithm was synthesized to examine the resource and power consumption, and it was observed that it requires $32$ clock cycles to encrypt the $64$-bit plain text with $80$-bit key. It occupies $1570$ gate equivalents (GE) which is suitable for the devices with limited resource. The number of rounds to encrypt or decrypt the data is $32$ which involves substitution and permutation of the plain text. The substitution and permutation network of the algorithm is shown in Figure \ref{fig: SP_Present}.

\subsection{KATAN}
\begin{figure}[ht]
	\begin{center}
		\includegraphics*[width=8cm]{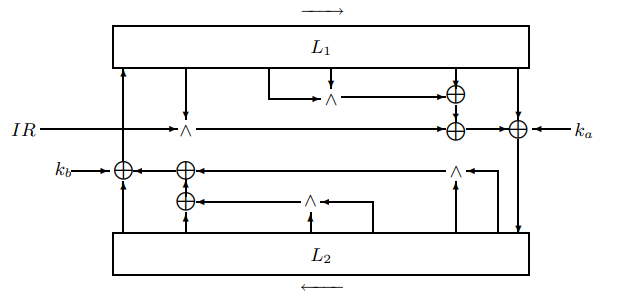}
	\end{center}
	\caption{Round of KATAN}
	\label{fig: KATAN}
\end{figure}

This cipher was developed for the resource constrained devices and consumes $48$\% less resources than PRESENT \cite{de2009katan}. The key scheduling is simple which is of $80$-bits and is generated using the Feistel structure to encrypt  $32$-bits, $48$-bits or $64$-bits of plain text. The shortcoming of the KATAN cipher is the number of rounds it takes to generate the cipher text. With $254$ rounds to generate on block of cipher text, the energy consumption is very high an throughput is low. On of the rounds of KATAN is shown in Figure \ref{fig: KATAN}

\subsection{Humming Bird}
Humming Bird was presented as an ultra-lightweight cryptographic algorithm for the resource constrained devices such as smart cards and RFID tags \cite{hummingbird}. It uses $256$-bit key to encrypt the $16$-bit block of plain text. The algorithm was implemented on a $16$-bit micro-controller and achieved better results in terms of throughput when compared to PRESENT. Since it uses separate functions for encryption and decryption, the resource footprint is larger. 

\subsection{SIMON and SPECK}
The SIMON and SPECK were proposed as a lightweight cryptography algorithms in \cite{simon}. They were developed by the National Security Agency (NSA) in the U.S. as an optimized algorithms for implementation in hardware and software. They require atleast $22$ rounds to encrypt the data and the number of mathematical operations are quite high.  

\subsection{RECTANGLE}
\begin{figure}[ht]
	\begin{center}
		\includegraphics*[width=8cm]{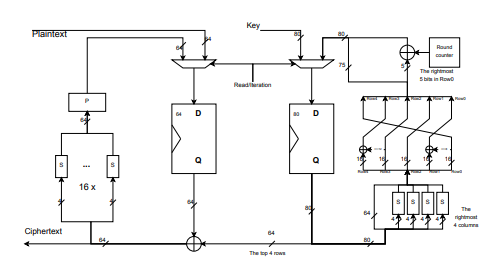}
	\end{center}
	\caption{Datapath of Rectangle-80}
	\label{fig: Rectangle}
\end{figure}

Bit-slice technique was utilized to make the algorithm lightweight and make the implementation faster. It uses the substitution and permutation networks to generate the cipher text in which the S-boxes are implemented in parallel whereas the permutation block uses $3$ rotations. For $80$-key, it requires $1600$ gates and which when implemented in parallel, yields a throughput of $246$ Kbits/sec at $100$ KHz clock. The datapath of the algorithm is depicted in Figure \ref{fig: Rectangle}
\subsection{SIT}
\begin{figure}[ht]
	\begin{center}
		\includegraphics*[width=8cm]{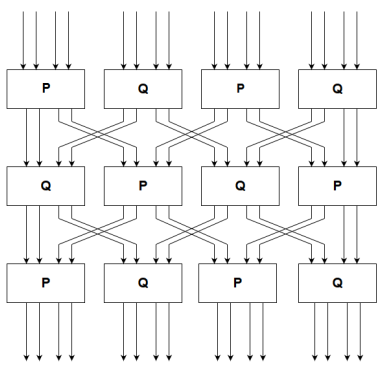}
	\end{center}
	\caption{F-function of SIT}
	\label{fig: f-function}
\end{figure}

A lightweight algorithm for the power constrained IoT devices named as SIT was proposed in \cite{Usman2017}. It is a combination of Feistel and SP network and therefore, can withstand the attacks on the resource constrained devices. It uses only $5$ rounds for encryption/decryption, and works on the block sizes of $64$-bits with key size of the same size. It is composed of two modules namely; key expansion and encryption, which are used for the key generation and data encryption respectively. The operations the key expansion involve XOR, concatenation, addition and shifting. The F-function shown in Figure \ref{fig: f-function} used in both key expansion and encryption/decryption block is inspired by the Khazad block cipher and performs the linear and non-linear transformation to remove any dependency of the output bits on the input bits \cite{barreto2000khazad}. This algorithm has been implemented on a low cost micro-controller on which it consumed only $22$ bytes of RAM, $3006$ and $2984$ clock cycles for encryption and decryption respectively. Furthermore, recently the FPGA implementation of the algorithm showed that a high throughput of upto $5682.56$ can be achieved which is $180$\% more than the LEA cipher \cite{mishrahigh}. 

\section{Conclusion}\label{sec: conclusion}
Security and privacy are the vital elements in the communication, and they are given more attention in the IoT because the data is usually human centered. The resource constrained devices in the IoT cannot utilize the state-of-the-art encryption algorithms, thus the development of lightweight encryption algorithm is indispensable.  In this correspondence the contemporary lightweight block ciphers have been studied. 

	\bibliographystyle{IEEEtran}
	\bibliography{Reference}
\end{document}